\documentclass[preprint,12pt,a4paper]{elsarticle}
\usepackage{amsmath,color,bm}
\usepackage{longtable}
\usepackage{bbold}
\usepackage{array}

\usepackage{todonotes}

\usepackage[utf8]{inputenc}
\usepackage{slashed}
\newcommand\Tstrut{\rule{0pt}{2.6ex}}         
\newcommand\Bstrut{\rule[-0.9ex]{0pt}{0pt}}   

\allowdisplaybreaks

\begin{document}

\begin{frontmatter}

  \title{                                                                                                              
     \vskip0.5cm
      Renormalization of three-quark operators with up to two derivatives at three loops
  }




  \author[mymainaddressa]{Bernd A. Kniehl}
      \ead{kniehl@desy.de}
      \address[mymainaddressa]{II.~Institut f\"ur Theoretische Physik, Universit\"at Hamburg,\\
        Luruper Chaussee 149, 22761 Hamburg, Germany}
      \author[mymainaddressb]{Oleg L. Veretin}
      \ead{oleg.veretin@desy.de}
      \address[mymainaddressb]{Institut f\"ur Theoretische Physik, Universit\"at Regensburg,\\
        Universit\"atsstrasse 31,93040 Regensburg, Germany}

      \begin{abstract}
       We study in QCD the $\overline{\mathrm{MS}}$ renormalization of three-quark operators with up to two covariant derivatives, which are related to $N=0,1,2$ Mellin moments of baryonic light-cone distributions amplitudes.   
        Apart from general three-quark operators, we also consider those corresponding to spin 3/2 and 1/2 states.
        We present in analytic form the renormalization constants and anomalous dimensions of these operators through three loops, confirming previous two- and three-loop results for $N=0$.
Furthermore, we evaluate through two loops their amputated four-point Green's functions with RI${}^\prime$/SMOM four-momentum assignment, which are required for the matching of lattice results with perturbative calculations.          
We work in linear covariant gauge and find the anomalous dimensions to be gauge independent as expected.
      \end{abstract}


\end{frontmatter}

\section{Introduction}

Distribution amplitudes (DAs) are fundamental non-perturbative functions describing the structure of nucleons~\cite{Efremov:1978cu,Lepage:1980fj,Chernyak:1983ej} in exclusive processes.
They are complementary to conventional parton distribution functions.
However, DA's are much less understood because their relationships to experimental observables are less direct.

The theoretical description of DAs is based on the relation of their Mellin moments to matrix elements of local composite operators.
Such matrix elements involve long-distance dynamics and, thus, cannot be accessed via perturbation theory alone.
The first two moments of the nucleon DA were estimated using QCD sum rules more than 30 years ago \cite{Chernyak:1984bm,King:1986wi,Chernyak:1987nu,Chernyak:1987nv}.

An alternative way to access the Mellin moments is to calculate them from first principles using lattice QCD.
Results for nucleon and hyperon DAs were presented in several papers (see, e.g., Ref.~\cite{Bali:2015ykx,RQCD:2019hps}).
They rely on computations of leading-twist baryonic matrix elements of local three-quark operators in the RI${}^\prime$/SMOM scheme \cite{Sturm:2009kb} with subsequent conversion to the $\overline{\mbox{MS}}$ renormalization scheme.
Recently, this analysis was updated in Ref.~\cite{Bali:2024oxg} using two-loop RI${}^\prime$/SMOM to $\overline{\mbox{MS}}$ conversion factors for the Mellin moments $N=0,1$ and evaluating their dependence on the $\overline{\mbox{MS}}$ renormalization scale $\mu$ using the corresponding anomalous dimensions at two loops.

As for perturbative-QCD ingredients for such lattice analyses, the state of the art is as follows.
The $\overline{\mbox{MS}}$ anomalous dimensions for $N=0$ were evaluated in Refs.~\cite{Pivovarov:1991nk,Krankl:2011gch} through two loops and in Refs.~\cite{Gracey:2012gx,Bali:2020isn} at three loops.
The cases $N=1,2$ were considered at one loop in Refs.~\cite{Bali:2015ykx,RQCD:2019hps,Bali:2020isn,Gruber:2017ozo}.
Our two-loop results for $N=1$ were presented in Ref.~\cite{Bali:2024oxg}.
The RI${}^\prime$/SMOM to $\overline{\mbox{MS}}$ conversion factors for $N=0,1,2$ were evaluated at one loop in Ref.~\cite{Gruber:2017ozo} and those for $N=0$ at two loops in Ref.~\cite{Kniehl:2022ido}.

The purpose of this paper is to push our knowledge of the three-loop anomalous dimensions to $N=1,2$ and of the two-loop conversion factors to $N=1$.
Doing this, we are faced with conceptual problems due to the mixing of evanescent operators with physical ones and the treatment of the Dirac matrix $\gamma_5$ in dimensional regularization.
In fact, this is related to the notion that the widely used $\overline{\mbox{MS}}$ prescription of subtracting poles in the regulator $\varepsilon$ of dimensional regularization in $d=4-2\varepsilon$ space-time dimensions does not uniquely fix a renormalization scheme.
A careful discussion of this may be found in Refs.~\cite{Krankl:2011gch,Gracey:2012gx}.
In this work, we take the advantage of the variant of $\overline{\mbox{MS}}$ scheme advocated in Ref.~\cite{Krankl:2011gch}, which exhibits the attractive features that evanescent operators are guaranteed to vanish in $d=4$ dimensions, so that one can work with physical (four-dimensional) operators only, and that $\gamma_5$ ambiguities are systematically avoided.
In Ref.~\cite{Krankl:2011gch}, the most general $N=0$ three-quark operator with open spinor indices was renormalized at two loops in this scheme.
This approach was tested for $N=0$ at three loops in Ref.~\cite{Gracey:2012gx} and at four loops in Ref.~\cite{Gracey:2025jnh}.
In this work, we apply it to higher Mellin moments, $N=1,2$, through three loops.

This paper is organized as follows.
In Section~2, we introduce the main notations and concepts.
In Section~3, we discuss the $\overline{\mbox{MS}}$ renormalization of the $N=0,1,2$ three-quark operators with open spinor indices and their counterparts for total spins 1/2 and 3/2.
In Section~4, we present the anomalous dimensions through three loops of the $N=0$ three-quark operator with open spinor indices and of their spin 1/2 and 3/2 counterparts with $N=0,1,2$.
In Section~5, we explain how we compute the RI${}^\prime$/SMOM to $\overline{\mbox{MS}}$ conversion factors for $N=1$ through two loops.
Section~6 contains our conclusions.
In the Appendix, we list the anomalous dimensions through three loops of the $N=1$ three-quark operators with open spinor indices.

All results of the calculations described in this paper, including also those not printed here, are provided in machine-readable form in an ancillary file published along with this paper.
This also includes the anomalous dimensions through three loops of the $N=2$ three-quark operators with open spinor indices, not listed in the Appendix.
As for the RI${}^\prime$/SMOM to $\overline{\mbox{MS}}$ conversion factors, we list both analytic expressions in terms of master integrals and numerical values with reasonable precision, more than sufficient for the matching with lattice results.

\section{Setup}

We start from the non-local three-quark operator with open Dirac spinor indices $\xi_i$ \cite{Braun:2000kw},
\begin{eqnarray}
  \label{nonlocal_op}
O_{\xi_1\xi_2\xi_3}&=&(\slashed{n}u^{c_1^\prime})_{\xi_1}(nx_1)[nx_1,nx_0]^{c_1^\prime c_1}(\slashed{n}d^{c_2^\prime})_{\xi_2}(nx_2)[nx_2,nx_0]^{c_2^\prime c_2}\nonumber\\
&&{}\times(\slashed{n}s^{c_3^\prime})_{\xi_3}(nx_3)[nx_3,nx_0]^{c_3^\prime c_3}\epsilon^{c_1c_2c_3}\,,
\end{eqnarray}
where $u,d,s$ are quark fields, $c_i$ are color indices in the fundamental representation of the SU(3) group, $n$ is an arbitrary dimensionless light-cone vector, with $n^2=0$, $x_i$ are scalar coordinates, and the gauge links,
\begin{equation}
[x,y]=P\exp\left[ig_s\int_0^1dt\,(x-y)\cdot A(tx+(1-t)y)\right]\,,
\end{equation}
with $P$ indicating the path-ordered product, $g_s$ being the strong-coupling constant, and $A$ being the gluon field, render Eq.~\eqref{nonlocal_op} gauge invariant.
The $\epsilon^{c_1c_2c_3}$ tensor endows Eq.~\eqref{nonlocal_op} with baryonic nature and neutralizes color.
Notice that Eq.~\eqref{nonlocal_op} is of leading twist by construction because higher-twist contributions come with more than one power of $n$ and turn out to be quenched by $n^2=0$.

To define baryon DAs, one needs to consider baryon-to-vacuum matrix elements of the type \cite{Braun:2000kw},
\begin{equation}
\label{matrix_elem}
\langle 0|O_{\xi_1\xi_2\xi_3}|B(p,\lambda)\rangle\,,
\end{equation}
where $B(p,\lambda)$ is a baryon state with four-momentum $p$ and helicity $\lambda$.
The spinor indices in Eq.~\eqref{nonlocal_op} can be contracted in different ways to obtain non-local baryonic currents with different quantum numbers.
Applying operator product expansion to Eq.~(\ref{nonlocal_op}), we can write it as a sum of
local operators multiplied by corresponding Wilson coefficients.
We refer to Ref.~\cite{Braun:2000kw} for the tensor decomposition and the definitions of the moments of DAs.

The $N=0$ Mellin moment of Eq.~(\ref{nonlocal_op}) with local-operator setting $x_1=x_2=x_3=x$ (off lightcone) was considered in great detail in Ref.~\cite{Gracey:2012gx}.
There, different local operators contributing to DAs were considered at three-loop order.
These operators can be classified according to irreducible representations of the Lorentz group.
We denote by $O^{(j\bar{j})}_{\pm}$ the operator that transforms according to the 
irreducible representation of the Lorentz group that is labeled by two Weyl-type spins $(j,\bar{j})$ and chirality $\pm$.
Two of these operators are of special interest \cite{Braun:1998id}, namely
\begin{eqnarray}
\label{op32}
O^{(\frac{3}{2},0)}_{+} &=&\slashed{n}u^{c_1^\prime}_L(x_1n)[x_1n,x_0n]^{c_1^\prime c_1}\slashed{n}d^{c_2^\prime}_L(x_2n)[x_2n,x_0n]^{c_2^\prime c_2} \nonumber\\
   &&{} \times\slashed{n}s^{c_3^\prime}_L(x_3n)[x_3n,x_0n]^{c_3^\prime c_3}\epsilon^{c_1c_2c_3}\,,\\
\label{op1}
O^{(1,\frac{1}{2})}_{-} &=&\slashed{n}u^{c_1^\prime}_L(x_1n)[x_1n,x_0n]^{c_1^\prime c_1}\slashed{n}d^{c_2^\prime}_L(x_2n)[x_2n,x_0n]^{c_2^\prime c_2}\nonumber\\
  &&{}\times\slashed{n}s^{c_3^\prime}_R(x_3n)[x_3n,x_0n]^{c_3^\prime c_3}\epsilon^{c_1c_2c_3}\,,
\end{eqnarray}
where we omit spinor indices for simplicity.
The left/right spinors are defined as
\begin{equation}
   q_{L,R}(x) = \frac{1 \mp \gamma_5}{2}\, q(x) \,.
\end{equation}
We do not consider operators of higher twist in this work.

The renormalization of three-quark operators requires special care. It is well known that the
$\overline{\text{MS}}$ prescription does not always fix a renormalization scheme completely.
The reason for that is that, in $d=4-2\varepsilon$ space-time dimensions, there exist infinitely many independent tensor structures, while there are only a finite number for $d=4$.
This effect naturally shows up in the case of four-fermion operators and in the case of three-quark operators under consideration here.
This implies that there exist operators in $d$ dimensions, which have no counterparts for $d=4$.
Examples of such operators are built using totally antisymmetric tensors of rank $n>4$.
Such structures vanish for $d=4$ and are conventionally called evanescent operators.
In the renormalization procedure, however, we have to take into account the mixing of physical and evanescent operators, since the latter yield finite contributions.
The mixing in the presence of evanescent operators was thoroughly analyzed in Refs.~\cite{Dugan:1990df,Herrlich:1994kh}. 

The complications with evanescent operators can be completely avoided by adopting the scheme introduced in Ref.~\cite{Krankl:2011gch}. Accordingly, we consider three-quark operators that have no contractions over spinor indices.
The renormalization of such an operator is plagued by a complicated mixing of the components of the corresponding operator multiplet and will be discussed in detail in the next section.
As an additional free bonus of this approach, we also completely get rid of the $\gamma_5$ problem of dimensional regularization, since no projections are taken and no Dirac traces containing $\gamma_5$ appear in such calculations.

There is, however, some complication in this approach, which does not yet show up for the lowest Mellin moment $N=0$, but manifests itself only for $N>0$.
The lowest-twist operators that can be constructed from $O_{\xi_1\xi_2\xi_3}$ are those of twist three.
In this case, the renormalization involves the mixing of $4\times4\times4=64$ components for $N=0$.
For $N>0$, there appear operators with covariant derivatives, which bring more indices and more components. The mixing pattern becomes more and more complicated, 
and the number of involved components grows rapidly with $N$ increasing, as $\sim64\times4^N$.
However, if we only consider operators of leading twist, then we have mixing of just $64\times(N+1)(N+2)/2$ components in the $N$th Mellin moment. Here, $(N+1)(N+2)/2$ is the number of ways to distribute $N$ covariant derivatives over three quarks.

\section{Renormalization of three-quark operators with open spinor indices}

Applying operator product expansion at the light cone to Eq.~(\ref{nonlocal_op}), 
we obtain local operators that are relevant for the $N$th Mellin moment of the DAs.
At leading twist, we have
\begin{eqnarray}
\label{local_op}
H_{\xi_1\xi_2\xi_3}^{opq}(x) &=& 
   \left[ (n\cdot D)^o (\slashed{n}u^{c_1})_{\xi_1}(x) \right] 
   \left[ (n\cdot D)^p (\slashed{n}d^{c_2})_{\xi_2}(x) \right]
\nonumber\\
    &&{}\times\left[ (n\cdot D)^q (\slashed{n}s^{c_3})_{\xi_3}(x) \right]\epsilon^{c_1c_2c_3}\,,
\end{eqnarray}
where $o,p,q=1,2,3\ldots$ with $N=o+p+q$, and $D$ is the covariant derivative.

To simplify the notation in the following formulae, we drop the spinor indices $\xi_i$ of the operators in Eq.~(\ref{local_op}).
In the first Mellin moment $N=0$, we just have a single operator, 
\begin{equation}
\label{op0}
O_{(0),1} = H^{000}\,.
\end{equation}
For $N=1$, we can build three operators,
\begin{equation}
\label{opN1}
\left(
  \begin{array}{c}
  O^{}_{(1),1}\\
  O^{}_{(1),2}\\
  O^{}_{(1),3}\\
  \end{array}
\right) =
\left(
  \begin{array}{c}
  H^{100}\\
  H^{010}\\
  H^{001}\\
  \end{array}
\right) \,.
\end{equation}
For $N=2$, we can build six operators distributing two derivatives in all possible ways,
\begin{equation}
\label{op2}
\left(
  \begin{array}{c}
  O_{(2),1}\\
  O_{(2),2}\\
  O_{(2),3}\\
  O_{(2),4}\\
  O_{(2),5}\\
  O_{(2),6}\\
  \end{array}
\right) =
\left(
  \begin{array}{c}
  H^{200}\\
  H^{020}\\
  H^{002}\\
  H^{110}\\
  H^{011}\\
  H^{101}\\
  \end{array}
\right)\,.
\end{equation}

Under renormalization, the operators with a given value of $N$ mix within the respective multiplets, (\ref{op0})--(\ref{op2}).
Specifically, we have
\begin{equation}
[O_{(N),k,\xi_1\xi_2\xi_3}] = \sum\limits_{k^\prime,\xi_1^\prime,\xi_2^\prime,\xi_3^\prime}
    Z_{(N),k,\xi_1\xi_2\xi_3}^{\phantom{(n)}k^\prime,\xi_1^\prime\xi_2^\prime\xi_3^\prime}
             O_{(N),k^\prime,\xi_1^\prime\xi_2^\prime\xi_3^\prime}\,,
\label{O_R}
\end{equation}
where $[O]$ is the renormalized version of operator $O$ and the $Z$ factors are renormalization constants.
It should be noted that Eq.~(\ref{O_R}) is valid only for the leading-twist operators (\ref{local_op}),
while the higher-twist operators also mix across different multiplets (\ref{op0})--(\ref{op2}).

To decompose the renormaliation constant $Z_{(N)}$, we introduce antisymmetric products of Dirac $\gamma$ matrices in $d$ dimensions,
\begin{eqnarray}
\Gamma_0 &=& \mathbb{1}\,, \nonumber\\
\Gamma_{\mu_1\mu_2} &=& \frac{1}{2!} \gamma_{[\mu_1} \gamma_{\mu_2]}\,, \nonumber\\
\Gamma_{\mu_1\mu_2\mu_3\mu_4} &=& \frac{1}{4!} \gamma_{[\mu_1} \gamma_{\mu_2} \gamma_{\mu_3} \gamma_{\mu_4]}\,,
    \qquad \mbox{etc.}\,,
\end{eqnarray}
where $[\cdots]$ implies total antisymmetrization.
Only products with even numbers of $\gamma$ matrices appear in our calculation.
In $d$ dimensions, the number of such structures is infinite.
However, for $d=4$, all products of more than four $\gamma$ matrices vanish,
\begin{equation}
  \Gamma_{\mu_1\dots\mu_n}\Big|_{d=4} =0 \qquad (n>4)  \,.
\end{equation}
In our three-loop calculation, we encounter only $\Gamma$ products of up to six $\gamma$ matrices.

We now decompose the operator matrix elements into tensor products of $\Gamma$ terms.
Altogether, we can build the following 24 structures:
\begin{eqnarray}
\Gamma_{nn0} &=& \Gamma_{\mu_1\dots\mu_n} \otimes \Gamma_{\mu_1\dots\mu_n} \otimes \Gamma_0
   \qquad (n=2,4,6) \,, \nonumber\\
\Gamma_{222} &=& \Gamma_{\mu_1\mu_2} \otimes \Gamma_{\mu_2\mu_3} \otimes \Gamma_{\mu_3\mu_1}\,,\nonumber\\
\Gamma_{422} &=& \Gamma_{\mu_1\mu_2\mu_3\mu_4} \otimes \Gamma_{\mu_1\mu_2} \otimes \Gamma_{\mu_3\mu_4}\,,\nonumber\\
\Gamma_{442} &=& \Gamma_{\mu_1\mu_2\mu_3\mu_4} \otimes \Gamma_{\mu_1\mu_2\mu_3\mu_5} 
          \otimes \Gamma_{\mu_4\mu_5}\,,\nonumber\\
\Gamma_{444} &=& \Gamma_{\mu_1\mu_2\mu_3\mu_4} \otimes \Gamma_{\mu_3\mu_4\mu_5\mu_6} 
          \otimes \Gamma_{\mu_5\mu_6\mu_1\mu_2}\,,\nonumber\\
\Gamma_{642} &=& \Gamma_{\mu_1\mu_2\mu_3\mu_4\mu_5\mu_6} 
         \otimes \Gamma_{\mu_1\mu_2\mu_3\mu_4} 
         \otimes \Gamma_{\mu_5\mu_6}\,, \nonumber\\
\Gamma_{624} &=& \Gamma_{\mu_1\mu_2\mu_3\mu_4\mu_5\mu_6} 
         \otimes \Gamma_{\mu_5\mu_6}
         \otimes \Gamma_{\mu_1\mu_2\mu_3\mu_4}  \,,
\label{Gdef2}
\end{eqnarray}
and the residual $\Gamma_{ijk}$ terms are obtained from Eq.~(\ref{Gdef2}) by cyclic permutations of $i,j,k$.
Notice that $\Gamma_{642}$ and $\Gamma_{624}$ are not related under such operation.
In Eq.~(\ref{Gdef2}), we have suppressed the spinor labels, always being $(\chi_i,\xi_i)$ for the $i$-th $\Gamma$ factor in each tensor product.

The renormalization group equation for the operator $[O_{(N)}]=Z_{(N)}O_{(N)}$ in Eq.~(\ref{O_R}) reads
\begin{equation}
  \left(\mu^2\frac{\partial}{\partial\mu^2}+\beta\frac{\partial}{\partial a}
  +\gamma_{(N)}\right)[O_{(N)}]=0\,,
\label{eq:rge}
\end{equation}
where $a=\alpha_s/(4\pi)$, with $\alpha_s=g_s^2/(4\pi)$ being the renormalized strong coupling, and
\begin{equation}
\mu^2\frac{d}{d\mu^2} a=\beta(a)=- b_0 a^2 - b_1 a^3 + \mathcal{O}(a^4)
\end{equation} 
is the Gell-Mann--Low function of QCD, with coefficients 
$b_0=11-2n_f/3$ \cite{Gross:1973id,Politzer:1973fx}, $b_1=102-38n_f/3$ \cite{Jones:1974mm,Caswell:1974gg}, {\it etc.}
Equation~\eqref{eq:rge} fixes the anomalous-dimension matrix order by order in perturbation theory,
\begin{equation}
\gamma_{(N)} = - \left(\mu^2\frac{d}{d\mu^2} Z_{(N)} \right) Z^{-1}_{(N)} =
  a \gamma_{(N)}^{(1)} + a^2 \gamma_{(N)}^{(2)} + a^3 \gamma_{(N)}^{(3 )} + \cdots\,.
\label{anom_def}
\end{equation}
Notice that the order of operators in Eq.~\eqref{anom_def} matters unless $N=0$.

Let us write the loop expansion of $Z_{(N)}$.
Omitting the subscript $(N)$, we have
\begin{equation}
  Z = 1 + \sum_{L=1}^\infty \sum_{K=1}^L \frac{a^L z_{LK}}{\varepsilon^K}\,.
\label{Zconst}
\end{equation}
At three-loop order, we need to invert $Z$ in Eq.~(\ref{Zconst}) only through order $\mathcal{O}(a^2)$, which yields
\begin{equation}
Z^{-1} = 1 - a\frac{z_{11}}{\varepsilon} + a^2\Big(
     \frac{-z_{22}+z_{11}^2}{\varepsilon^2} - \frac{z_{21}}{\varepsilon} \Big) \,.
\label{invZconst}  
\end{equation}
Inserting Eqs.~(\ref{Zconst}) and (\ref{invZconst}) into Eq.~(\ref{anom_def}) and expanding through three loops, we obtain
\begin{eqnarray}
\gamma^{(1)} &=& z_{11}\,,\nonumber\\
\gamma^{(2)} &=& 2z_{21} + \frac{1}{\varepsilon}( 2 z_{22} - z_{11}^2 + b_0 z_{11})\,,\nonumber\\
\gamma^{(3)} &=& 3z_{31}
  + \frac{1}{\varepsilon}( 3 z_{32} 
    - 2 z_{21} z_{11} - z_{11} z_{21} + 2b_0 z_{21} + b_1 z_{11}) \nonumber\\
    &&{}+ \frac{1}{\varepsilon^2}[ 3 z_{33} 
      - 2z_{22} z_{11} - z_{11} z_{22} + z_{11}^3 + b_0(2z_{22} - z_{11}^2)]\,.
\end{eqnarray}
Notice that $z_{LK}$ do not in general commute with each other, and renormalizability requires that all $\gamma^{(k)}$ terms are finite in the limit $\varepsilon\to0$.

\section{Anomalous dimensions}

We evaluate the renormalization constants $Z_{(N)}$ and anomalous dimensions $\gamma_{(N)}$ for the non-local three-quark operators with $N=0,1,2$ covariant derivatives through three-loop order, both for the case of open spinor indices as in Eq.~(\ref{nonlocal_op}) and for the cases with spin and chirality assignments as in Eqs.~(\ref{op32}) and (\ref{op1}).
For the case of Eq.~(\ref{nonlocal_op}), we give $\gamma_{(0)}$ below and $\gamma_{(1)}$ in the Appendix.
For the cases of Eqs.~(\ref{op32}) and (\ref{op1}), we list $\gamma_{(N)}$ for $N=0,1,2$ further below.

To evaluate the pure ultraviolet divergent parts of Feynman diagrams and to separate them from the infrared divergent ones, we use the method of global infrared rearrangement \cite{Misiak:1994zw,Chetyrkin:1997fm}.
This allows us to reduce the expressions to massive tadpole diagrams.

The form of $\gamma_{(0)}$ for Eq.~(\ref{nonlocal_op}) is particularly simple and reads:
\begin{eqnarray}
\label{gamma_N0_1}
\gamma^{(1)}_{(0)} &=&-\frac{1}{3} \mathbb{C}_2\,,\\
\gamma^{(2)}_{(0)} &=& (70-4n_f) \mathbb{C}_0 
  + \left( -\frac{49}{18} + \frac{1}{27}n_f \right) \mathbb{C}_2
  + \frac{1}{9} \mathbb{C}_4
  + \frac{1}{18} \mathbb{C}_{42}\,,\\
\gamma^{(3)}_{(0)} &=& \left( \frac{16492}{9} - \frac{1708}{9} n_f + \frac{20}{9} n_f^2 - \frac{434}{3} \zeta_3 \right) \mathbb{C}_0
  \nonumber\\
&&{}+ \left( -\frac{5479}{108} + \frac{212}{81} n_f + \frac{13}{81} n_f^2 + \frac{127}{9} \zeta_3 + 
  \frac{40}{9} n_f \zeta_3 \right) \mathbb{C}_2
  \nonumber\\
&&{}+ \left( -\frac{127}{216} - \frac{2}{81} n_f - \frac{13}{36} \zeta_3 \right) \mathbb{C}_4
  + \left( -\frac{19}{27} - \frac{1}{162} n_f - \frac{25}{18} \zeta_3 \right) \mathbb{C}_{42}
  \nonumber\\
&&{}+ \left( \frac{101}{1296} - \frac{59}{432} \zeta_3 \right) \mathbb{C}_6
    + \left( \frac{1}{16} - \frac{1}{16} \zeta_3 \right) \mathbb{C}_{642}\,,
\label{gamma_N0_3}
\end{eqnarray}
where we have introduced the symmetric combinations
\begin{eqnarray}
\mathbb{C}_0 &=& \Gamma_{000}\,, \nonumber\\
\mathbb{C}_2 &=& \Gamma_{022} + \Gamma_{202} + \Gamma_{220}\,, \nonumber\\
\mathbb{C}_4 &=& \Gamma_{044} + \Gamma_{404} + \Gamma_{440}\,, \nonumber\\
\mathbb{C}_{42} &=& \Gamma_{422} + \Gamma_{242} + \Gamma_{224}\,, \nonumber\\
\mathbb{C}_6 &=& \Gamma_{066} + \Gamma_{606} + \Gamma_{660}\,, \nonumber\\
\mathbb{C}_{642} &=& \Gamma_{642} + \Gamma_{426} + \Gamma_{264} + \Gamma_{246} + \Gamma_{462} + \Gamma_{624}\,.
\end{eqnarray}
Notice that, in Eq.~(\ref{gamma_N0_3}), we have also exposed the evanescent structures $\mathbb{C}_6$ and $\mathbb{C}_{642}$, which vanish for $d=4$.

We can now compare these results at two loops with Ref.~\cite{Krankl:2011gch} and at three loops with Ref.~\cite{Gracey:2012gx}.
To this end, we observe that the results of Refs.~\cite{Krankl:2011gch,Gracey:2012gx} are presented in different bases. Specifically, the structures $\mathbb{C}_{42}$, $\mathbb{C}_6$, and $\mathbb{C}_{642}$ 
are rewritten in favor of $\mathbb{C}_2^2$, $\mathbb{C}_2^3$, and $\mathbb{C}_2\mathbb{C}_4$ using the following $d$-dimensional identities:
\begin{eqnarray}
\mathbb{C}_{42} &=&- 3 d(d-1) \mathbb{C}_0 - 2 (d-3) \mathbb{C}_2 - \frac{1}{2} \mathbb{C}_4 +
\frac{1}{2} \mathbb{C}_2^2 \,,\nonumber\\
\mathbb{C}_6 &=&- 12 d(d-1)(2d-1) \mathbb{C}_0 - 3 (d-1)(7d-24) \mathbb{C}_2 - 6 (2d-5) \mathbb{C}_4
\nonumber \\
 &&{}+ 2 (3d-4) \mathbb{C}_2^2 - \frac{1}{2} \mathbb{C}_2^3 + \frac{3}{2} \mathbb{C}_2 \mathbb{C}_4 +
3 \mathbb{C}_{444} \,,\nonumber\\
\mathbb{C}_{642} &=& 12 d(d-1)(2d-7) \mathbb{C}_0 + 9 (d^2-9d+16) \mathbb{C}_2 + 2 (2d-5) \mathbb{C}_4
\nonumber \\
&&{}- 2 (3d-10) \mathbb{C}_2^2 + \frac{1}{2} \mathbb{C}_2^3 -
\frac{1}{2} \mathbb{C}_2 \mathbb{C}_4 - 3 \mathbb{C}_{444}\,.
\label{id3}
\end{eqnarray}
In Refs.~\cite{Gracey:2012gx,Krankl:2011gch}, Eq.~(\ref{id3}) was applied before renormalization, which effectively leads to finite renormalizations of anomalous dimensions. 
Taking this into account, we find full agreement with Refs.~\cite{Krankl:2011gch,Gracey:2012gx}.

Proceeding to higher Mellin moments, $N=1,2$, we find that the anomalous dimensions are no longer expressible in terms of symmetric combinations $\mathbb{C}_X$, but rather in terms of all $\Gamma_{ijk}$ structures.
In fact, one could appropriately generalize Eq.~(\ref{id3}) to such cases, rewriting the newly appearing higher Dirac tensors in terms of squares and cubes of lower ones like $\Gamma_{022}$.
However, since there are no obvious reasons in favor of one or another representation, we minimize intermediate transformations and represent our result in terms of structures linear in $\Gamma_{ijk}$.

We now turn to the two operators in Eqs.~(\ref{op32}) and (\ref{op1}) with $N=0,1,2$ covariant derivatives.
To evaluate their anomalous dimensions, we need to know the actions of $\Gamma_{ijk}$ on them.
It is straightforward to compute the following commutators in $d$ dimensions:
\begin{align}
[\Gamma_{220},\Gamma_{202}] &= -8 \Gamma_{222} \,,\label{comm1} \\
[\Gamma_{440},\Gamma_{220}]&= 0\,,\label{comm2} \\
[\Gamma_{440},\Gamma_{202}] &= -16\Gamma_{442} \,,\\
[\Gamma_{422}, \Gamma_{022}] &= 8\Gamma_{424} - 8\Gamma_{442} \,,\\
[\Gamma_{422}, \Gamma_{202}] &= -8\Gamma_{424} - 16(d-3)\Gamma_{222} \,, \label{comm5}\\
[\Gamma_{222}, \Gamma_{202}] &= 4\Gamma_{422} - 4\Gamma_{242} 
    + 4(d-2)\Gamma_{202} - 4(d-2)\Gamma_{022} \,,
\end{align}
and cyclic permutations.

From Eq.~(\ref{comm1}), we conclude that
\begin{equation}
\Gamma_{222} \,O^{(\frac{3}{2},0)}_+ = \Gamma_{222} \, O^{(1,\frac{1}{2})}_- = 0 \,.
\end{equation} 
Moreover, for $d=4$, we have $\Gamma_{442}=\Gamma_{424}=\Gamma_{244}=0$, and it follows from Eqs.~(\ref{comm2})--(\ref{comm5}) that both $O^{(\frac{3}{2},0)}_-$ and $O^{(1,\frac{1}{2})}_-$ are eigenvectors of the following ten tensors: $\{\Gamma_{022}$, $\Gamma_{202}$, $\Gamma_{220}$, $\Gamma_{222}$, $\Gamma_{044}$, $\Gamma_{404}$, $\Gamma_{440}$, $\Gamma_{422}$, $\Gamma_{242}$, $\Gamma_{224}\}$.
The respective eigenvalues are summarized in Table~1.

\begin{table}[h]
\begin{tabular}{| c|ccc|c|ccc|ccc|}
\hline
 $O$ \Tstrut\Bstrut 
       & $\Gamma_{022}$ & $\Gamma_{202}$ & $\Gamma_{220}$ 
       & $\Gamma_{222}$ 
       & $\Gamma_{044}$ & $\Gamma_{404}$ & $\Gamma_{440}$ 
       & $\Gamma_{422}$ & $\Gamma_{242}$ & $\Gamma_{224}$ \\
\hline
 $O^{(\frac{3}{2},\,0)}_{+}$ \Tstrut\Bstrut 
       & $-4$ & $-4$ & $-4$ 
       & 0
       & 24 & 24 & 24 
       &  8 &  8 & 8 \\
 $O^{(1,\,\frac{1}{2})}_{-}$ \Tstrut\Bstrut 
       & 0 & 0 & $-4$
       & 0
       & $-24$ & $-24$ & 24
       &  0 &  0 &  $-8$ \\
\hline
\end{tabular}
\caption{Eigenvalues of the $\Gamma_{ijk}$ tensors for the eigenvectors $O^{(\frac{3}{2},0)}_+$ and $O^{(1,\frac{1}{2})}_-$.}
\end{table}

Using the eigenvalues from Table~1 , we obtain the anomalous dimensions of the operators 
in Eq.~(\ref{op32}) and (\ref{op1}) for $N=0,1,2$.

For $N=0$, we have
\begin{eqnarray}
\gamma^{(\frac{3}{2},0)}_{(0),+} &=& 4 a + a^2 \left( \frac{328}{3} - \frac{40}{9} n_f \right)
\nonumber\\
  &&{}+ a^3\left( 2382 - \frac{2008}{9} n_f + \frac{8}{27} n_f^2 - \frac{1120}{3} \zeta_3 
- \frac{160}{3} n_f \zeta_3 \right) \,,\nonumber\\
\gamma^{(1,\frac{1}{2})}_{(0),-} &=& \frac{4}{3} a + a^2 \left( \frac{236}{3} - \frac{112}{27} n_f \right)
\nonumber\\
  &&{}+ a^3\left( \frac{18496}{9} - \frac{16168}{81} n_f + \frac{128}{81} n_f^2 - \frac{544}{3} \zeta_3 
- \frac{160}{9} n_f \zeta_3 \right) \,.
\end{eqnarray}

For $N=1$, we have
\begin{eqnarray}
a_{11} &=& a_{22}=a_{33}
 = a \left( \frac{20}{3} \right)
+ a^2 \left(
  \frac{1184}{9} - \frac{184 n_f}{27}
  \right)
\nonumber\\
&&{}+ a^3 \left(
   \frac{3254219}{1215} - \frac{185686 n_f}{729} - \frac{200 n_f^2}{243} 
      - \frac{11096 \zeta_3}{45} - \frac{800 n_f \zeta_3}{9}
  \right)\,,
\nonumber\\
a_{12} &=& a_{13}=a_{21}=a_{23}=a_{31}=a_{32}
=a \left( -\frac{4}{3} \right)
+ a^2 \left(
   -\frac{100}{9} + \frac{32 n_f}{27}
  \right)
\nonumber\\
&&{}+ a^3 \left(
   -\frac{360089}{2430} + \frac{11519 n_f}{729} + \frac{136 n_f^2}{243} 
      - \frac{2852 \zeta_3}{45} + \frac{160 n_f \zeta_3}{9} 
  \right)\, .
\end{eqnarray}
and $\gamma_{(1),-}^{(1,\frac{1}{2})}=(b_{ij})$ with
\begin{eqnarray}
b_{11} &=& b_{22}
  = a \left( \frac{40}{9} \right)
+ a^2 \left(
  \frac{26744}{243} - \frac{524 n_f}{81} 
  \right)
\nonumber\\
&&{}+ a^3 \left(
   \frac{156285211}{65610} - \frac{513052 n_f}{2187} + \frac{188 n_f^2}{729} 
       - \frac{9598 \zeta_3}{135} - \frac{1600 n_f \zeta_3}{27}
  \right)\,,
\nonumber\\
b_{33} &=&
a \left( \frac{44}{9} \right)
+ a^2 \left(
    \frac{8986}{81} - \frac{520 n_f}{81}
  \right)
\nonumber\\
&&{}+ a^3 \left(
   \frac{54536053}{21870} - \frac{172702 n_f}{729} + \frac{40 n_f^2}{729} 
     - \frac{17042 \zeta_3}{135} - \frac{1760 n_f \zeta_3}{27}
  \right)\,,
\nonumber\\
b_{12} &=& b_{21}
=a \left( -\frac{4}{3} \right)
+ a^2 \left(
    -\frac{3530}{243} + \frac{32 n_f}{27}
  \right)
\nonumber\\
&&{}+ a^3 \left(
    -\frac{4087478}{32805} + \frac{36097 n_f}{2187} + \frac{136 n_f^2}{243} 
      - \frac{10402 \zeta_3}{135} + \frac{160 n_f \zeta_3}{9} 
  \right)\,,
\nonumber\\
b_{13} &=& b_{23}=b_{31}=b_{32}
=a \left( -\frac{16}{9} \right)
+ a^2 \left(
    -\frac{1414}{81} + \frac{92 n_f}{81}
  \right)
\nonumber\\
&&{}+ a^3 \left(
    -\frac{100789}{486} + \frac{1517 n_f}{81} + \frac{556 n_f^2}{729} 
      - \frac{896 \zeta_3}{27} + \frac{640 n_f \zeta_3}{27} 
  \right)\,.
\nonumber\\
\end{eqnarray}

To present our results for $N=2$, it is convenient to switch to the notation of Eq.~(\ref{anom_def}).
We have
\begin{equation}
\gamma_{(2),+}^{(\frac{3}{2},0),(1)} =
\left(
  \begin{array}{*6{>{\scriptstyle}c}}
  \frac{76}{9}   & -\frac{4}{9} & -\frac{4}{9} &  -\frac{8}{3} &  0 &  -\frac{8}{3} \\
  -\frac{4}{9} & \frac{76}{9}   & -\frac{4}{9} & -\frac{8}{3} & -\frac{8}{3} & 0   \\
  -\frac{4}{9} & -\frac{4}{9} &  \frac{76}{9}  &  0     &    -\frac{8}{3}& -\frac{8}{3}\\
  -\frac{8}{9} & -\frac{8}{9} &  0    & \frac{28}{3}  & -\frac{4}{3} & -\frac{4}{3} \\
  0  &  -\frac{8}{9} & -\frac{8}{9} & -\frac{4}{3} & \frac{28}{3} & -\frac{4}{3} \\
  -\frac{8}{9} & 0 & -\frac{8}{9} & -\frac{4}{3} & -\frac{4}{3} & \frac{28}{3} \\
  \end{array}
\right)\,,
\end{equation}
\begin{equation}
\gamma_{(2),+}^{(\frac{3}{2},0),(2)} =
\left(
  \begin{array}{*6{>{\scriptscriptstyle}c}}
  \frac{12062}{81} - \frac{692}{81}n_f & -\frac{607}{243} + \frac{26}{81}n_f & 
    -\frac{607}{243} + \frac{26}{81}n_f & -\frac{619}{27} + \frac{62}{27} & 
      \frac{64}{81} & -\frac{619}{27} + \frac{62}{27}n_f \\
  -\frac{607}{243} + \frac{26}{81}n_f &  \frac{12062}{81} - \frac{692}{81}n_f & -\frac{607}{243} + \frac{26}{81}n_f &
      -\frac{619}{27} + \frac{62}{27}n_f & -\frac{619}{27} + \frac{62}{27}n_f & \frac{64}{81} \\
  -\frac{607}{243} + \frac{26}{81}n_f & -\frac{607}{243} + \frac{26}{81}n_f &  \frac{12062}{81} - \frac{692}{81}n_f & 
    \frac{64}{81} & -\frac{619}{27} + \frac{62}{27}n_f & -\frac{619}{27} + \frac{62}{27}n_f \\
  -\frac{703}{81} + \frac{70}{81}n_f & -\frac{703}{81} + \frac{70}{81}n_f & \frac{16}{243} & 
    \frac{12545}{81} - \frac{82}{9}n_f & -\frac{932}{81} + \frac{32}{27}n_f & -\frac{932}{81} + \frac{32}{27}n_f \\
  \frac{16}{243} & -\frac{703}{81} + \frac{70}{81}n_f & -\frac{703}{81} + \frac{70}{81}n_f & 
    -\frac{932}{81} + \frac{32}{27}n_f &  \frac{12545}{81} - \frac{82}{9}n_f & -\frac{932}{81} + \frac{32}{27}n_f \\
   -\frac{703}{81} + \frac{70}{81}n_f & \frac{16}{243} & -\frac{703}{81} + \frac{70}{81}n_f & 
     -\frac{932}{81} + \frac{32}{27}n_f & -\frac{932}{81} + \frac{32}{27}n_f & \frac{12545}{81} - \frac{82}{9}n_f \\
  \end{array}
\right)\,,
\end{equation}
\begin{eqnarray}
\gamma_{(2),+}^{(\frac{3}{2},0),(3)} &=& (c_{ij}) \,,\nonumber\\
c_{11} &=& c_{22} = c_{33}\nonumber\\
&=&    \frac{383453249}{131220} - \frac{406331}{1458}n_f - \frac{1042}{729}n_f^2 
    -\frac{74968}{405}\zeta_3 - \frac{3040}{27}\zeta_3 n_f 
    \,,\nonumber\\
    c_{12} &=& c_{13}  = c_{21} = c_{23} = c_{31} = c_{32}\nonumber\\
  &=&-\frac{3396733}{87480} + \frac{31967}{8748}n_f + \frac{187}{729}n_f^2 
   - \frac{8996}{405}\zeta_3 + \frac{160}{27}\zeta_3 n_f
   \,,\nonumber\\
   c_{44} &=& c_{55} = c_{66}\nonumber\\
   &=&   \frac{87854419}{29160} - \frac{825397}{2916}n_f - \frac{467}{243}n_f^2 
    - \frac{22952}{135}\zeta_3 - \frac{1120}{9}\zeta_3 n_f
    \,,\nonumber\\
    c_{43} &=& c_{45}  = c_{54} = c_{56} = c_{64} = c_{65}
    \nonumber\\
  &=&   -\frac{1869902}{10935} + \frac{11738}{729}n_f + \frac{136}{243}n_f^2 - \frac{716}{15}\zeta_3 
   + \frac{160}{9}\zeta_3 n_f
   \,,\nonumber\\
c_{14} &=& c_{16}  = c_{24} = c_{25} = c_{35} = c_{36}\nonumber\\
&=&    -\frac{27263477}{87480} + \frac{27259}{972}n_f + \frac{89}{81}n_f^2 
     - \frac{12448}{135}\zeta_3 + \frac{320}{9}\zeta_3 n_f
     \,,\nonumber\\
c_{15} &=& c_{26}  = c_{34}\nonumber\\
  &=&   \frac{499003}{10935} - \frac{146}{243}n_f - \frac{1408}{45}\zeta_3
   \,,\nonumber\\
c_{41} &=& c_{42}  = c_{52} = c_{53} = c_{61} = c_{63}\nonumber\\
   &=&-\frac{31997597}{262440} + \frac{11653}{972}n_f + \frac{221}{729}n_f^2 
   - \frac{12448}{405}\zeta_3 + \frac{320}{27}\zeta_3 n_f
   \,,\nonumber\\
c_{46} &=& c_{51}  = c_{62}\nonumber\\
  &=&-\frac{1869902}{10935} + \frac{11738}{729}n_f + \frac{136}{243}n_f^2 - \frac{716}{15}\zeta_3 
   + \frac{160}{9}\zeta_3 n_f
   \,,
\end{eqnarray}
\begin{equation}
\gamma_{(2),-}^{(1,\frac{1}{2}),(1)} =
\left(
  \begin{array}{*6{>{\scriptstyle}c}}
  \frac{58}{9} & -\frac{4}{9} & -\frac{2}{3} & -\frac{8}{3} & 0 & -\frac{10}{3} \\
  - \frac{4}{9} & \frac{58}{9} & -\frac{2}{3} & -\frac{8}{3} & -\frac{10}{3} & 0 \\
  -\frac{2}{3} & -\frac{2}{3} & \frac{64}{9} & 0 & -\frac{10}{3} & -\frac{10}{3} \\
  -\frac{8}{9} & -\frac{8}{9} & 0 & \frac{68}{9} & -\frac{16}{9} & -\frac{16}{9} \\
   0 & -\frac{10}{9} & -\frac{10}{9} & -\frac{16}{9} & \frac{70}{9} & -\frac{4}{3} \\
  -\frac{10}{9} & 0 &-\frac{10}{9} & -\frac{16}{9} & -\frac{4}{3} & \frac{70}{9} \\
  \end{array}
\right)\,,
\end{equation}
\begin{eqnarray}
\lefteqn{\gamma_{(2),-}^{(1,\frac{1}{2}),(2)}}\nonumber\\
&=&
\left(
  \begin{array}{*6{>{\scriptscriptstyle}c}}
  \frac{31865}{243} - \frac{665 n_f}{81} & -\frac{1043}{243} + \frac{26 n_f}{81} & -\frac{497}{81} 
   + \frac{7 n_f}{27} & -\frac{1477}{54} + \frac{62 n_f}{27} & 77/162 & -\frac{4961}{162} + \frac{61 n_f}{27} \\
 -\frac{1043}{243} + \frac{26 n_f}{81} & 
  \frac{61927}{486} - \frac{665 n_f}{81} & -\frac{739}{162} + \frac{7 n_f}{27} & -\frac{1477}{54} 
    + \frac{62 n_f}{27} & -\frac{2239}{81} + \frac{61 n_f}{27} &  \frac{77}{162} \\
 -\frac{17}{3} + \frac{7 n_f}{27} & -\frac{221}{54} + \frac{7 n_f}{27} & 
  \frac{64313}{486} - \frac{662 n_f}{81} & \frac{2}{3} & -\frac{4325}{162} + \frac{61 n_f}{27} &
     -\frac{2404}{81} + \frac{61 n_f}{27} \\
  -\frac{181}{18} + \frac{70 n_f}{81} & -\frac{181}{18} + \frac{70 n_f}{81} & -\frac{22}{243} & 
  \frac{22265}{162} - \frac{706 n_f}{81} & -\frac{481}{27} + \frac{92 n_f}{81} & -\frac{1229}{81} 
    + \frac{92 n_f}{81} \\
  -\frac{29}{162} & -\frac{806}{81} + \frac{71 n_f}{81} & -\frac{4939}{486} + \frac{71 n_f}{81} &
    -\frac{1393}{81} + \frac{92 n_f}{81} & \frac{33548}{243} - \frac{707 n_f}{81} &
      -\frac{7117}{486} + \frac{32 n_f}{27} \\
  -\frac{595}{54} + \frac{71 n_f}{81} & -\frac{29}{162} & -\frac{2729}{243} + \frac{71 n_f}{81} & 
    -\frac{131}{9} + \frac{92 n_f}{81} & -\frac{7117}{486} + \frac{32 n_f}{27} & 
       \frac{67261}{486} - \frac{707 n_f}{81} \\
  \end{array}
\right)\,,\
\end{eqnarray}
\begin{eqnarray}
\gamma_{(2),-}^{(1,\frac{1}{2}),(3)} &=& (d_{ij}) \,,\nonumber\\
d_{11} &=&
\frac{173894021}{65610} - \frac{2283025 n_f}{8748} - \frac{671 n_f^2}{1458} - 
\frac{4018 \zeta_3}{135} - \frac{2320 n_f \zeta_3}{27} 
\,,\nonumber\\
d_{12} &=&
-\frac{910847}{65610} + \frac{8107 n_f}{2187} + \frac{187 n_f^2}{729} - \frac{5044 \zeta_3}{135} + 
\frac{160 n_f \zeta_3}{27}
\,,\nonumber\\
d_{13} &=&
-\frac{5727761}{87480} + \frac{4343 n_f}{972} + \frac{169 n_f^2}{486} - \frac{116 \zeta_3}{15} + 
\frac{80 n_f \zeta_3}{9}
\,,\nonumber\\
d_{14} &=&
-\frac{5926451}{19440} + \frac{28691 n_f}{972} + \frac{89 n_f^2}{81} - \frac{3751 \zeta_3}{45} + 
\frac{320 n_f \zeta_3}{9}
\,,\nonumber\\
d_{15} &=&
\frac{7100327}{174960} - \frac{1619 n_f}{2916} - \frac{1607 \zeta_3}{45} 
\,,\nonumber\\
d_{16} &=&
-\frac{69481721}{174960} + \frac{24107 n_f}{729} + \frac{229 n_f^2}{162} - 
  \frac{1829 \zeta_3}{45} + \frac{400 n_f \zeta_3}{9}
\,,\nonumber\\
d_{21} &=&
-\frac{1197857}{65610} + \frac{8107 n_f}{2187} + \frac{187 n_f^2}{729} - 
     \frac{4804 \zeta_3}{135} + \frac{160 n_f \zeta_3}{27}
\,,\nonumber\\
d_{22} &=&
\frac{701684249}{262440} - \frac{2279269 n_f}{8748} - \frac{671 n_f^2}{1458} - 
\frac{4798 \zeta_3}{135} - \frac{2320 n_f \zeta_3}{27}
\,,\nonumber\\
d_{23} &=&
-\frac{972878}{10935} + \frac{4451 n_f}{972} + \frac{169 n_f^2}{486} + \frac{1514 \zeta_3}{135} + 
     \frac{80 n_f \zeta_3}{9}
\,,\nonumber\\
d_{24} &=&
-\frac{663581}{2160} + \frac{28691 n_f}{972} + \frac{89 n_f^2}{81} - \frac{3724 \zeta_3}{45} + 
\frac{320 n_f \zeta_3}{9}
\,,\nonumber\\
d_{25} &=&
-\frac{72086837}{174960} + \frac{94853 n_f}{2916} + \frac{229 n_f^2}{162} - 
     \frac{866 \zeta_3}{15} + \frac{400 n_f \zeta_3}{9}
\,,\nonumber\\
d_{26} &=&
\frac{8621489}{174960} - \frac{1619 n_f}{2916} - \frac{1874 \zeta_3}{45}
\,,\nonumber\\
d_{31} &=&
-\frac{5746271}{87480} + \frac{12685 n_f}{2916} + \frac{169 n_f^2}{486} - 
     \frac{644 \zeta_3}{135} + \frac{80 n_f \zeta_3}{9}
\,,\nonumber\\
d_{32} &=&
-\frac{3709427}{43740} + \frac{13009 n_f}{2916} + \frac{169 n_f^2}{486} + \frac{62 \zeta_3}{5} + 
     \frac{80 n_f \zeta_3}{9}
\,,\nonumber\\
d_{33} &=&
\frac{146838325}{52488} - \frac{1157353 n_f}{4374} - \frac{565 n_f^2}{729} - 
\frac{2468 \zeta_3}{27} - \frac{2560 n_f \zeta_3}{27}
\,,\nonumber\\
d_{34} &=&
\frac{608918}{10935} - \frac{715 n_f}{1458} - \frac{1883 \zeta_3}{45}
\,,\nonumber\\
d_{35} &=&
-\frac{4345316}{10935} + \frac{23326 n_f}{729} + \frac{229 n_f^2}{162} - \frac{2749 \zeta_3}{45} + 
     \frac{400 n_f \zeta_3}{9}
\,,\nonumber\\
d_{36} &=&
-\frac{33665971}{87480} + \frac{94879 n_f}{2916} + \frac{229 n_f^2}{162} -\frac{217 \zeta_3}{5} + 
     \frac{400 n_f \zeta_3}{9}
\,,\nonumber\\
d_{41} &=&
-\frac{21070183}{174960} + \frac{18263 n_f}{1458} + \frac{221 n_f^2}{729} - 
\frac{3671 \zeta_3}{135} + \frac{320 n_f \zeta_3}{27}
\,,\nonumber\\
d_{42} &=&
-\frac{21245317}{174960} + \frac{18263 n_f}{1458} + \frac{221 n_f^2}{729} - 
     \frac{3524 \zeta_3}{135} + \frac{320 n_f \zeta_3}{27}
\,,\nonumber\\
d_{43} &=&
\frac{1300517}{131220} + \frac{1213 n_f}{4374} - \frac{1603 \zeta_3}{135}
\,,\nonumber\\
d_{44} &=&
\frac{1519157}{540} - \frac{194039 n_f}{729} - \frac{761 n_f^2}{729} - \frac{8654 \zeta_3}{135} - 
     \frac{2720 n_f \zeta_3}{27}
\,,\nonumber\\
d_{45} &=&
-\frac{39922753}{174960} + \frac{55367 n_f}{2916} + \frac{556 n_f^2}{729} - 
     \frac{2296 \zeta_3}{135} + \frac{640 n_f \zeta_3}{27}
\,,\nonumber\\
d_{46} &=&
-\frac{4978391}{19440} + \frac{18301 n_f}{972} + \frac{556 n_f^2}{729} - \frac{37 \zeta_3}{15} + 
     \frac{640 n_f \zeta_3}{27}
\,,\nonumber\\
d_{51} &=&
\frac{6667997}{524880} + \frac{397 n_f}{1458} - \frac{1927 \zeta_3}{135}
\,,\nonumber\\
d_{52} &=&
-\frac{78328067}{524880} + \frac{39625 n_f}{2916} + \frac{605 n_f^2}{1458} - 
     \frac{2798 \zeta_3}{135} + \frac{400 n_f \zeta_3}{27}
\,,\nonumber\\
d_{53} &=&
-\frac{19957589}{131220} + \frac{119959 n_f}{8748} + \frac{605 n_f^2}{1458} - 
     \frac{2869 \zeta_3}{135} + \frac{400 n_f \zeta_3}{27}
\,,\nonumber\\
d_{54} &=&
-\frac{40063783}{174960} + \frac{54607 n_f}{2916} + \frac{556 n_f^2}{729} - 
     \frac{1696 \zeta_3}{135} + \frac{640 n_f \zeta_3}{27}
\,,\nonumber\\
d_{55} &=&
\frac{738003409}{262440} - \frac{584758 n_f}{2187} - \frac{1685 n_f^2}{1458} - 
     \frac{3988 \zeta_3}{135} - \frac{2800 n_f \zeta_3}{27}
\,,\nonumber\\
d_{56} &=& 
-\frac{7889975}{52488} + \frac{36745 n_f}{2187} + \frac{136 n_f^2}{243} - 
     \frac{1481 \zeta_3}{27} + \frac{160 n_f \zeta_3}{9}
\,,\nonumber\\
d_{61} &=&
-\frac{77659931}{524880} + \frac{13471 n_f}{972} + \frac{605 n_f^2}{1458} - 
     \frac{1909 \zeta_3}{135} + \frac{400 n_f \zeta_3}{27}
\,,\nonumber\\
d_{62} &=&
\frac{5623079}{524880} + \frac{397 n_f}{1458} - \frac{1834 \zeta_3}{135}
\,,\nonumber\\
d_{63} &=&
-\frac{39843811}{262440} + \frac{122323 n_f}{8748} + \frac{605 n_f^2}{1458} - 
     \frac{611 \zeta_3}{45} + \frac{400 n_f \zeta_3}{27}
\,,\nonumber\\
d_{64} &=&
-\frac{1626227}{6480} + \frac{54143 n_f}{2916} + \frac{556 n_f^2}{729} -\frac{ 31 \zeta_3}{45} + 
     \frac{640 n_f \zeta_3}{27}
\,,\nonumber\\
d_{65} &=&
-\frac{7578449}{52488} + \frac{36745 n_f}{2187} + \frac{136 n_f^2}{243} - 
     \frac{1553 \zeta_3}{27} + \frac{160 n_f \zeta_3}{9}
\,,\nonumber\\
d_{66} &=&
\frac{185536411}{65610} - \frac{2342365 n_f}{8748} - \frac{1685 n_f^2}{1458} - 
\frac{8258 \zeta_3}{135} - \frac{2800 n_f \zeta_3}{27}
\,.
\end{eqnarray}

\section{Matrix elements of $N=1$ three-quark operators with open spinor indices}

\begin{figure}[h]
\centerline{\includegraphics[width=0.3\textwidth,viewport=0 0 211 217,clip]{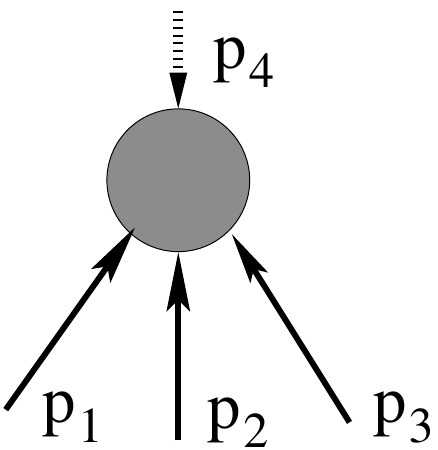}}
\caption{\label{fig:one}%
  Matrix element $\langle{\cal O}(p_4)\bar{u}(p_1)\bar{d}(p_2)\bar{s}(p_3)\rangle$ of a three-quark operator in momentum space, where we omit all spinor and color indices.
The four-momentum $p_4=-(p_1+p_2+p_3)$ is the one coming into the operator.}
\end{figure}

In order to exploit QCD results on baryonic quantities from lattice simulations for perturbative calculations in the continuum, we need to evaluate matrix elements of three-quark operators in kinematic configurations as implemented on the lattice order by order in perturbation theory using a renormalization scheme appropriate for the continuum.
Here, we do this for amputated four-point Green's functions involving three-quark operators of Lorentz spin $N=1$ with RI${}^\prime$/SMOM kimematics using $\overline{\mathrm{MS}}$ renormalization.
Since the kinematic configuration is fixed in momentum space, we are led to consider the Fourier transform \cite{Bali:2024oxg},
\begin{eqnarray}
  H_{\chi_1\chi_2\chi_3,\xi_1\xi_2\xi_3}^{opq}(p_1,p_2,p_3)
  &=&-\int\mathrm{d}^4x_1\mathrm{d}^4x_2\mathrm{d}^4x_3\,\mathrm{e}^{\mathrm{i}(p_1\cdot x_1+p_2\cdot x_2+p_3\cdot x_3)}\nonumber\\
  &&{}\times \langle H_{\chi_1\chi_2\chi_3}^{opq}(0)\bar{u}_{\xi_1^\prime}^{c_1}(x_1)
  \bar{d}_{\xi_2^\prime}^{c_2}(x_2)
  \bar{s}_{\xi_3^\prime}^{c_3}(x_3)\rangle\epsilon^{c_1c_2c_3}\nonumber\\
  &&{}\times G_{2,\xi_1^\prime\xi_1}^{-1}(p_1)
  G_{2,\xi_2^\prime\xi_2}^{-1}(p_2)
  G_{2,\xi_3^\prime\xi_3}^{-1}(p_3)\,,
  \label{amp}
\end{eqnarray}
where $\langle\cdots\rangle$ indicates the functional integral with the gauge fields taken in a specific gauge and $p_1,p_2,p_3$ are incoming four-momenta assigned to the three external quark legs (see Fig.~\ref{fig:one}).
The two-point function $G_2(p)$ required for the amputation of the external quark legs is defined by
\begin{equation}
G_{2,\xi\xi^\prime}(p)\delta^{cc^\prime}=\int\mathrm{d}^4x\,\mathrm{e}^{\mathrm{i}p\cdot x}
\langle u_{\xi}^c(0)\bar{u}_{\xi^\prime}^{c^\prime}(x)\rangle\,.  
\end{equation}
By four-momentum conservation, the inserted three-quark operator carries incoming four-momentum $p_4=-p_1-p_2-p_3$.
We adopt RI${}^\prime$/SMOM kinematics from Refs.~\cite{Bali:2015ykx,RQCD:2019hps,Bali:2024oxg,Kniehl:2022ido}, with
\begin{align}
\label{kin}
p_1^2 = p_2^2 = p_3^2 = p_4^2 = -\mu_E^2\,, \qquad
p_1\cdot p_2 = p_1\cdot p_3 = \frac{\mu_E^2}{2}\,, \qquad p_2\cdot p_3=0\,,
\end{align}
where $\mu_E$ is a Euclidean subtraction point, i.e.\ we have $\mu_E^2>0$ in Minkowski metric with signature $(+,-,-,-,)$.
This implies that $p_1\cdot p_4=0$ and $p_2\cdot p_4=p_3\cdot p_4=\mu_E^2/2$.
The fully symmetric choice $p_i^2=-\mu_E^2$ and $p_i\cdot p_j=\mu_E^2/3$ for $i\neq j$ is inconvenient to implement on the lattice.
We work in linear covariant gauge, keeping the gauge parameter generic.

In Ref.~\cite{Kniehl:2022ido}, the amplitude in Eq.~(\ref{amp}) was evaluated at two loops in QCD for general three-quark operators with $N=0$, keeping spinor indices open.
In the present work, we extend these results to the case of $N=1$ at leading twist.
The computation of finite contributions in RI${}^\prime$/SMOM kinematics being rather tedious, we refrain from reaching beyond two loops for the time being.
We adhere to the procedure of Ref.~\cite{Kniehl:2022ido}, which we briefly summarize, and only outline the modifications required by the case of $N=1$.

For the ease of notation, we introduce the collective spinor label $\bm{\chi}=(\chi_1,\chi_2,\chi_3,\xi_1,\xi_2,\xi_3)$.
In the case of $N=0$, where we just have the single operator in Eq.~\eqref{op0}, we also omit the labels $o,p,q$.
According to Ref.~\cite{Kniehl:2022ido}, the amplitude in Eq.~(\ref{amp}) can be decomposed as
\begin{align}
\label{amp_decomposition}
H_{\bm{\chi}}(p_1,p_2,p_3)&=\sum_{j=1}^{N_d}T_{j,\bm{\chi}}(p_1,p_2,p_3)f_j\,,
\end{align}
where $T_{j,\bm{\chi}}(p_1,p_2,p_3)$ are tensor structures and $f_j$ are scalar form factors.
The number $N_d$ of linearly independent tensor structures depends on the loop order and on whether $d=4$ or $d=4-2\varepsilon$ is considered.
At one loop (two loops), we have $N_4=64$ (247) and $N_{d\neq4}=67$ (581), $N_{d\neq4}-N_4=3$ (247) tensor structures being evanescent \cite{Kniehl:2022ido}.
The latter mix with the four-dimensional tensor structures under renormalization and must be taken into account \cite{Kniehl:2022ido}.
In principle, $f_j$ can be extracted from Eq.~\eqref{amp_decomposition} by applying suitably constructed projectors.
However, the latter depend non-trivially on $d$.
In Ref.~\cite{Kniehl:2022ido}, we thus advocated a different procedure, namely to first renormalize the left-hand side of Eq.~\eqref{amp_decomposition} in the $\overline{\mathrm{MS}}$ scheme and then to put $d=4$.
This results in an over-determined, but correct system of equations for $f_j$, which we solve using standard linear algebra avoiding the use of projectors.
At one loop (two loops), this yields 67 (581) form factors $f_j$, 64 (247) of which are independent.
We note that, although $T_{j,\bm{\chi}}$ do depend on the specific four-momenta $p_1,p_2,p_3$, $f_j$ only depend on their scalar products in Eq.~\eqref{kin}, i.e., on the single energy scale $\mu_E$.
Choosing $\mu=\mu_E$ as we do, $f_j$ become just numbers.
Their $\mu$ dependencies may easily be retrieved by solving the renormalization group equation in Eq.~\eqref{eq:rge}, which yields
\begin{eqnarray}
  f_j(\mu)&=&c_j^{(0)}+a\left(c_j^{(1)}-\gamma_{(0)}^{(1)}c_j^{(0)}\ell\right)
  +a^2\left\{c_j^{(2)}-\left[\gamma_{(0)}^{(2)}c_j^{(0)}+\left(\gamma_{(0)}^{(1)}-b_0\right)c_j^{(1)}\right]\ell
  \phantom{\frac{\gamma_{(0)}^{(1)}}{2}}\right.
  \nonumber\\
  &&{}+\left.\frac{\gamma_{(0)}^{(1)}}{2}\left(\gamma_{(0)}^{(1)}+b_0\right)c_j^{(0)}\ell^2\right\}
  +\mathcal{O}(a^3)\,,
\label{eq:run}
\end{eqnarray}
where $c_j^{(L)}=f_j^{(L)}(\mu_E)$ and $\ell=\ln(\mu^2/\mu_E^2)$.
In Ref.~\cite{Kniehl:2022ido}, all 67 (581) one-loop (two-loop) form factors $f_j$ are expressed in terms of 4 (44) master integrals, which are evaluated analytically (numerically).

Turning to the case of $N=1$, we have the triplet of operators in Eq.~\eqref{opN1}, with label $k=1,2,3$.
Furthermore, we have one Lorentz index $\alpha$ from the covariant derivative, which is saturated by $n$ in Eq.~\eqref{local_op}.
It turns out that the tensor decomposition in Eq.~(\ref{amp_decomposition}) requires just minimal modifications in the step from $N=0$ to $N=1$.
In fact, we can write
\begin{align}
\label{amp_decompositionmod}
H_{k,\bm{\chi}}^{\alpha}(p_1,p_2,p_3)&=\sum_{j=1}^{N_d}p_k^{\alpha}T_{j,\bm{\chi}}(p_1,p_2,p_3)f_{k,j},\qquad (k=1,2,3)\,,
\end{align}
with the very same tensor structures $T_{j,\bm{\chi}}$ as in Eq.~\eqref{amp_decomposition}.
Therefore, we can exploit the decomposition developed in Ref.~\cite{Kniehl:2022ido} upon obvious modifications to evaluate all form factors $f_{k,j}$, being again numbers for $\mu=\mu_E$, whose $\mu$ dependence may be retrieved from Eq.~\eqref{eq:run} with the replacements $f_j\to\vec{f}_j=(f_{1,j},f_{2,j},f_{3,j})^T$ and $\gamma_{(0)}^{(L)}\to\gamma_{(1)}^{(L)}$ being $3\times3$ matrices.
For general value of $N$, $\vec{f}_j=(f_{1,j},\ldots,f_{k_{\mathrm{max}},j})^T$ and $\gamma_{(N)}^{(L)}$ are $k_{\mathrm{max}}\times k_{\mathrm{max}}$ matrices with $k_{\mathrm{max}}=(N+1)(N+2)/2$.

Our calculational procedure is similar to Refs.~\cite{Kniehl:2022ido,Kniehl:2020sgo,Kniehl:2020nhw}.
We generate the contributing Feynman diagrams using the computer program QGRAF \cite{Nogueira:1991ex}.
We project out and evaluate the color and spinor traces using custom-made routines written in FORM \cite{Vermaseren:2000nd} language.
We reduce the resulting Feynman integrals to a small set of master integrals using the Laporta algorithm \cite{Laporta:2000dsw} of integration by parts \cite{Chetyrkin:1981qh} as implemented in FIRE \cite{Smirnov:2014hma}. 
Besides the integration-by-parts relations, we exploit additional relations arising from the symmetric kinematics to further reduce the number of master integrals, down to 2 one-loop master integrals, 
to be evaluated analytically, and 44 two-loop master integrals, to be evaluated numerically.
For the latter, we apply sector decomposition \cite{Binoth:2000ps,Binoth:2003ak} using FIESTA \cite{Smirnov:2015mct} and CUBA \cite{Hahn:2004fe}.
At two-loops, we reach a relative numerical precision of about $10^{-6}$ for the individual master integrals.
In the ancillary file published in the journal along with this paper, we list $f_j$ and $f_{k,j}$ both as numerical values and as analytic expressions in terms of master integrals, allowing for more precise numerical or even analytic evaluations in the future.\footnote{We caution the reader that the face values of $f_j$ for the case of $N=0$ differ from those presented in the ancillary file published along with Ref.~\cite{Kniehl:2022ido}.
This reflects the ambiguity in the special solution of an inhomogeneous system of linear equations.
This ambiguity disappears in the evaluation of $H_{\bm{\chi}}(p_1,p_2,p_3)$ in Eq.~\eqref{amp_decomposition}.}

The leading poles in $\varepsilon$ that occur in the evaluation Eq.~\eqref{amp} for $N=1$ prior to renormalization allow us to crosscheck our results for the respective anomalous dimensions, $\gamma_{(1)}^{(L)}$ with $L=1,2$.

\section{Conclusions}
\label{sec:five}

In this work, we considered the light-cone operator product expansion of non-local three-quark operators and evaluated the renormalization constants and anomalous dimensions of their Mellin moments $N=0,1,2$ through three loops in the $\overline{\mathrm{MS}}$ scheme.
Furthermore, for $N=1$, we computed the respective four-point Green's functions with RI${}'$/SMOM four-momentum assignments \cite{Sturm:2009kb} frequently adopted in lattice QCD simulations \cite{RQCD:2019hps,Bali:2024oxg}, with all external legs having non-vanishing invariant masses, through two loops.
This extends our preceding work \cite{Kniehl:2022ido}, where the case of $N=0$ was considered.
The new results allows one to extract from lattice QCD the first two Mellin moments, with $N=0,1$, of baryonic DAs at two loops with three-loop evolution.

We took advantage of the approach developed in Ref.~\cite{Krankl:2011gch} to avoid complications due to renormalization mixing with evanescent operators and the treatment of the $\gamma_5$ Dirac matrix in dimensional regularization.
In turn, we had to pay the price of dealing with open spinor indices, which rendered bookkeeping somewhat cumbersome.

We considerably pushed the state of the art beyond previous achievements \cite{Bali:2024oxg,Gracey:2012gx,Kniehl:2022ido}.
The pioneering three-loop analysis of Ref.~\cite{Gracey:2012gx} adopted a less infrared-safe RI${}'$/SMOM subtraction point, involving one external leg with zero invariant mass, and was confined to $N=0$.
The two-loop analysis of Ref.~\cite{Kniehl:2022ido} adopted the favorable RI${}'$/SMOM subtraction point \cite{RQCD:2019hps,Bali:2024oxg}, but was also confined to $N=0$.
In Ref.~\cite{Bali:2024oxg}, the anomalous dimensions for $N=1$ were presented through two loops.

\bigskip

\section*{Acknowledgments}

We are grateful to Vladimir M. Braun, Meinulf G\"ockeler, and Alexander N. Manashov for fruitful discussions.
O.L.V. is grateful to the University of Hamburg for the warm hospitality.
This work was supported in part by the German Research Foundation DFG through
Research Unit FOR~2926 ``Next Generation Perturbative QCD for Hadron Structure:
Preparing for the Electron-Ion Collider'' under Grant Nos.~KN~365/13-2 and 409651613.


\appendix

\section{Anomalous dimensions of $N=1$ three-quark operators with open spinor indices}

In Eq.~(\ref{gamma_N0_1})--(\ref{gamma_N0_3}), we presented the anomalous dimension $\gamma_{(0)}$ of the $N=0$ three-quark operator with open spinor indices in Eq.~(\ref{nonlocal_op}) through three loops.
Here, we list the corresponding $N=1$ Mellin moment, $\gamma_{(1)}$.

It is convenient to decompose the $L$-loop coefficients of $\gamma_{(1)}$ as 
\begin{equation}
\gamma_{(1)}^{(L)} = \sum\limits_{X\in X_L} M_X \otimes \Gamma_X\,, 
\end{equation}
where, at $L$ loops, index $X$ takes values from list $X_L$,
\begin{eqnarray}
X_1 &=& \{000, 022, 202, 220\} \,, \nonumber\\
X_2 &=& X_1 \cup \{044, 404, 440, 422, 242, 224, 222 \} \,, \nonumber\\
X_3 &=& X_2 \cup \{244, 424, 442, 444, 066, 606, 660, 642, 426, 264,
246, 462, 624\}\,,\nonumber\\
&&
\end{eqnarray}
and $M_X$ are $3\times3$ matrices corresponding to the operator basis in Eq.~(\ref{opN1}).
Thanks to the underlying cyclic symmetry of baryonic nature, it is sufficient to consider only the following sets of indices:
\begin{equation}
\{000, 022, 044, 422, 222, 244, 444, 066, 642, 246\}\,.
\end{equation}
The residual matrices, pertaining to cyclic permutations of indices $lmn$, emerge by respective shifts of rows and columns.
Explicitly, this means that 
\begin{eqnarray}
||M_{nlm}||_{ij} &=& ||M_{lmn}||_{i^-j^-}\,, \nonumber\\
||M_{mnl}||_{ij} &=& ||M_{lmn}||_{i^+j^+}\,, \nonumber
\end{eqnarray}
where $i^-$ ($i^+$) implies that row index $i$ is decreased (increased) by one in cyclic sense and similarly for the column index $j$.

At the one-loop level, we have
\begin{eqnarray}
M_{000} &=&
  \begin{pmatrix}
      \frac{32}{9} &  - \frac{16}{9} &  - \frac{16}{9} \\
    - \frac{16}{9} &    \frac{32}{9} &  - \frac{16}{9} \\
    - \frac{16}{9} &  - \frac{16}{9} &    \frac{32}{9} \\
  \end{pmatrix}\,,\nonumber
\\
M_{022} &=&
  \begin{pmatrix}
    - \frac{1}{3} &  0 &  0 \\
    0 &  - \frac{2}{9} &  - \frac{1}{9} \\
    0 &  - \frac{1}{9} &  - \frac{2}{9} \\
  \end{pmatrix}\,.
\end{eqnarray}

At the two-loop level, we have
\begin{eqnarray}
M_{000} &=&
  \begin{pmatrix}
  \frac{8348}{81} - \frac{508}{81} n_f  &  
       - \frac{1339}{81} + \frac{92}{81} n_f &  - \frac{1339}{81} + \frac{92}{81} n_f \\
  - \frac{1339}{81} + \frac{92}{81} n_f & 
       \frac{8348}{81} - \frac{508}{81} n_f &  - \frac{1339}{81} + \frac{92}{81} n_f \\
  - \frac{1339}{81} + \frac{92}{81} n_f &  
      - \frac{1339}{81} + \frac{92}{81} n_f & \frac{8348}{81} - \frac{508}{81} n_f \\
  \end{pmatrix}\,,\nonumber
  \\
M_{022} &=&
  \begin{pmatrix}
   - \frac{1265}{486} + \frac{1}{27} n_f &  - \frac{55}{243} &,  - \frac{55}{243} \\
   - \frac{29}{486} &  - \frac{817}{486} + \frac{4}{81} n_f &  - \frac{22}{27} - \frac{1}{81} n_f \\
   - \frac{29}{486} &  - \frac{22}{27} - \frac{1}{81} n_f &  - \frac{817}{486} + \frac{4}{81} n_f \\
  \end{pmatrix}\,,\nonumber
\\
M_{044} &=&
  \begin{pmatrix}
  \frac{1}{9} &  0 &  0 \\
  0 & \frac{109}{1944} & \frac{107}{1944} \\
  0 & \frac{107}{1944} & \frac{109}{1944} \\
  \end{pmatrix}\,,\nonumber
\\
M_{422} &=&
  \begin{pmatrix}
  - \frac{4}{243}  & - \frac{5}{486} &  - \frac{5}{486} \\
  - \frac{19}{972} & - \frac{1}{27}  &  - \frac{2}{243} \\
  - \frac{19}{972} & - \frac{2}{243} &  - \frac{1}{27}  \\
  \end{pmatrix}\,,\nonumber
\\
M_{222} &=&
  \begin{pmatrix}
    0 & \frac{2}{81} &  - \frac{2}{81} \\
   - \frac{2}{81} &  0 & \frac{2}{81} \\
    \frac{2}{81} &  - \frac{2}{81} &  0 \\
  \end{pmatrix} \,.
\end{eqnarray}

At the three-loop level, we have
\begin{eqnarray}
M_{000} &=& 
   \begin{pmatrix}
   a & b & b \\
   b & a & b \\
   b & b & a \\
  \end{pmatrix}\,,  
\qquad\text{with} \nonumber\\
a &=& \frac{99412663}{43740} - \frac{28673}{405} \zeta_3 
    + \Big(- \frac{165626}{729} - \frac{1280}{27} \zeta_3 \Big) n_f + \frac{508}{729} n_f^2 \,,
\nonumber\\
b &=& - \frac{19261543}{87480} - \frac{29917}{810} \zeta_3 
    + \Big( \frac{13639}{729} + \frac{640}{27} \zeta_3 \Big) n_f + \frac{556}{729} n_f^2\,,\nonumber 
\end{eqnarray}
\begin{eqnarray}
M_{022} &=& 
   \begin{pmatrix}
   a & c & c \\
   d & b & e \\
   d & e & b \\
  \end{pmatrix}\,,
\qquad\text{with} \nonumber\\
a &=& - \frac{150115}{2916} + \frac{464}{27} \zeta_3 
  + \Big( \frac{5615}{2187} + \frac{40}{9} \zeta_3 \Big) n_f + \frac{13}{81} n_f^2 \,,
\nonumber\\
b &=& - \frac{14893789}{524880} + \frac{10693}{1620} \zeta_3 
  + \Big( \frac{8561}{4374} + \frac{80}{27} \zeta_3 \Big) n_f + \frac{80}{729} n_f^2 \,, 
\nonumber\\
c &=& - \frac{2441}{1458} + \frac{176}{2187} n_f - \frac{83}{54} \zeta_3 \,,
\nonumber\\
d &=& \frac{1091}{2916} + \frac{109}{4374} n_f - \frac{83}{54} \zeta_3 \,,
\nonumber\\
e &=& - \frac{10855391}{524880} + \frac{14657}{1620} \zeta_3 
  + \Big( \frac{845}{1458} + \frac{40}{27} \zeta_3 \Big) n_f + \frac{37}{729} n_f^2 \,,\nonumber
\end{eqnarray}
\begin{eqnarray}
M_{044} &=& 
   \begin{pmatrix}
   a & c & c \\
   d & b & e \\
   d & e & b \\
  \end{pmatrix}\,,
\qquad\text{with} \nonumber\\
a &=& - \frac{120977}{209952} - \frac{199}{648} \zeta_3 - \frac{2}{81} n_f\,, \nonumber\\
b &=& - \frac{8089}{58320} - \frac{439}{810} \zeta_3 - \frac{187}{8748} n_f\,, \nonumber\\
c &=& \frac{9629}{419904} - \frac{35}{1296} \zeta_3\,, \nonumber\\
d &=& - \frac{2467}{419904} - \frac{35}{1296} \zeta_3\,, \nonumber\\
e &=& - \frac{991381}{2099520} + \frac{449}{2160} \zeta_3 - \frac{29}{8748} n_f\,,\nonumber 
\end{eqnarray}
\begin{eqnarray}
M_{422} &=& 
   \begin{pmatrix}
   a & c & c \\
   d & b & e \\
   d & e & b \\
  \end{pmatrix}\,,
\qquad\text{with} \nonumber\\
a &=& - \frac{7943}{58320} - \frac{1201}{1620} \zeta_3 + \frac{8}{2187} n_f\,, \nonumber\\
b &=& - \frac{379687}{1049760} - \frac{1007}{1080} \zeta_3\,, \nonumber\\
c &=& - \frac{27757}{116640} - \frac{1049}{3240} \zeta_3 - \frac{4}{2187} n_f\,, \nonumber\\
d &=& - \frac{33097}{116640} - \frac{1049}{3240} \zeta_3 - \frac{43}{8748} n_f\,, \nonumber\\
e &=& - \frac{5461}{52488} - \frac{43}{324} \zeta_3 - \frac{19}{4374} n_f\,,\nonumber
\end{eqnarray}
\begin{eqnarray}
M_{222} &=& 
   \begin{pmatrix}
   0 & a & -a \\
   -a & 0 & a \\
   a & -a & 0 \\
  \end{pmatrix}\,,
\qquad\text{with} \nonumber\\
a &=& \frac{6901}{6480} - \frac{13}{729} n_f - \frac{1139}{540} \zeta_3\,,\nonumber
\end{eqnarray}
\begin{eqnarray}
M_{244} &=&
\left(
  \begin{array}{*3{>{\scriptstyle}c}}
  0 &  - \frac{50759}{233280} + \frac{767}{6480} \zeta_3 & \frac{50759}{233280} - \frac{767}{6480} \zeta_3 \\
  - \frac{689}{25920} + \frac{11}{2160} \zeta_3 &
             \frac{47}{729} - \frac{5}{81} \zeta_3 &  - \frac{35719}{233280} + \frac{367}{6480} \zeta_3 \\
  \frac{689}{25920} - \frac{11}{2160} \zeta_3 &
            \frac{35719}{233280} - \frac{367}{6480} \zeta_3 &  - \frac{47}{729} + \frac{5}{81} \zeta_3 \\
  \end{array}
\right)\,,\nonumber
\\
M_{444} &=&
\left(
  \begin{array}{*3{>{\scriptstyle}c}}
  \frac{7231}{209952} & \frac{9293}{419904} & \frac{9293}{419904} \\
  \frac{9293}{419904} & \frac{7231}{209952} & \frac{9293}{419904} \\
  \frac{9293}{419904} & \frac{9293}{419904} & \frac{7231}{209952} \\
  \end{array}
  \right)\,,\nonumber 
\\
M_{066} &=&
\left(
  \begin{array}{*3{>{\scriptstyle}c}}
  \frac{101}{1296} - \frac{59}{432} \zeta_3 &  0 &  0 \\
  0 & \frac{10343}{419904} - \frac{67}{1296} \zeta_3 & \frac{22381}{419904} - \frac{55}{648} \zeta_3 \\
  0 & \frac{22381}{419904} - \frac{55}{648} \zeta_3 & \frac{10343}{419904} - \frac{67}{1296} \zeta_3 \\
  \end{array}
\right)\,,\nonumber
\\
M_{642} &=&
\left(
  \begin{array}{*3{>{\scriptstyle}c}}
  \frac{3905}{209952} - \frac{1}{48} \zeta_3 &
      \frac{8807}{279936} - \frac{1}{32} \zeta_3 & \frac{6127}{839808} - \frac{1}{96} \zeta_3 \\
  \frac{6239}{279936} - \frac{61}{2592} \zeta_3 &
       \frac{13037}{419904} - \frac{37}{1296} \zeta_3 & \frac{11369}{839808} - \frac{1}{96} \zeta_3 \\
  \frac{18151}{839808} - \frac{47}{2592} \zeta_3 &
      - \frac{7}{839808} - \frac{7}{2592} \zeta_3 & \frac{1}{24} - \frac{1}{24} \zeta_3 \\
  \end{array}
  \right)\,,\nonumber
\\
M_{246} &=&
\left(
  \begin{array}{*3{>{\scriptstyle}c}}
  \frac{1}{24} - \frac{1}{24} \zeta_3 &
       - \frac{7}{839808} - \frac{7}{2592} \zeta_3 & \frac{18151}{839808} - \frac{47}{2592} \zeta_3 \\
  \frac{11369}{839808} - \frac{1}{96} \zeta_3 &
        \frac{13037}{419904} - \frac{37}{1296} \zeta_3 & \frac{6239}{279936} - \frac{61}{2592} \zeta_3 \\
  \frac{6127}{839808} - \frac{1}{96} \zeta_3 &
       \frac{8807}{279936} - \frac{1}{32} \zeta_3 & \frac{3905}{209952} - \frac{1}{48} \zeta_3 \\
  \end{array}
  \right)\,.
\end{eqnarray}

\end{document}